\documentclass[manuscript]{acmart}

\AtBeginDocument{%
  \providecommand\BibTeX{{%
    \normalfont B\kern-0.5em{\scshape i\kern-0.25em b}\kern-0.8em\TeX}}}
 
 \acmConference {ACM Knowledge Discovery and Data Mining (KDD) 2020 Conference Workshop "Fragile Earth: Data Science for a Sustainable Planet"}{August 23--27, 2020}{Virtual Conference}

\begin{document}

\title{Leveraging traditional ecological knowledge in ecosystem restoration projects utilizing machine learning}

\author{Bogdana Rakova}
\affiliation{
  \institution{Accenture, Responsible AI}
  \streetaddress{415 Mission St Floor 35}
  \city{San Francisco}
  \country{USA}}
  
\email{bogdana.rakova@accenture.com}


\author{Alexander Winter}
\affiliation{
  \institution{California Institute of Integral Studies}
  \streetaddress{1453 Mission St}
  \city{San Francisco}
  \country{USA}}
\email{alwinter@mymail.ciis.edu}


\begin{abstract}
Ecosystem restoration has been recognized to be critical to achieving accelerating progress on all of the United Nations' Sustainable Development Goals. Decision makers, policymakers, data scientists, earth scientists, and other scholars working on these projects could positively benefit from the explicit consideration and inclusion of diverse perspectives. Community engagement throughout the stages of ecosystem restoration projects could contribute to improved community well-being, the conservation of biodiversity, ecosystem functions, and the resilience of socio-ecological systems. Conceptual frameworks are needed for the meaningful integration of traditional ecological knowledge of indigenous peoples and local communities with data science and machine learning work practices. Adaptive frameworks would consider and address the needs and challenges of local communities and geographic locations by improving community and inter-agent communication around restoration and conservation projects and by making relevant real-time data accessible. In this paper, we provide a brief analysis of existing Machine Learning (ML) applications for forest ecosystem restoration projects. We go on to question if their inherent limitations may prevent them from being able to adequately address socio-cultural aspects of the well-being of all involved stakeholders. Bias and unintended consequences pose significant risks of downstream negative implications of ML-based solutions. We suggest that adaptive and scalable practices could incentivize interdisciplinary collaboration during all stages of ecosystemic ML restoration projects and align incentives between human and algorithmic actors. Furthermore, framing ML projects as open and reiterative processes can facilitate access on various levels and create incentives that lead to catalytic cooperation in the scaling of restoration efforts.

\end{abstract}


\maketitle

\section{Introduction}

United Nations (UN) has recently announced 2021-2030 to be the UN decade on ecosystem restoration \cite{un_strategy} and aims to act as an accelerator towards the achievement of a shared vision: "A world where we have restored the relationship between humans and nature: Where we increase the area of healthy ecosystems and put a stop to their loss and degradation – for the health and well-being of all life on earth and that of future generations" \cite{un_partnership}.

Ecosystem restoration holds increasing potential for maintaining and increasing the rates of carbon sequestration \cite{projectdrawdown}. Furthermore, it conserves biodiversity through maintaining a range of ecosystem services and safeguarding rich cultures and traditional ways of life. Indigenous and local knowledge systems for ecosystem restoration have been recognized by researchers as key components of sustainable development \cite{berkes1993traditional}. Reyes-Garcia V et al. (2019) propose that actively involving indigenous peoples and local communities (IPLCs) in restoration efforts (1) can help in selecting sites and species, (2) can increase local participation in the planning, execution, and monitoring of restoration activities, and (3) can provide historical information on ecosystem state and management \cite{reyes2019contributions}. 

The importance of IPLCs knowledge systems for environmental sustainability have been acknowledged by the UN System Task Team on the Post 2015 UN Development Agenda, stating that "traditional and indigenous knowledge, adaptation, and coping strategies can be major assets for local response strategies" \cite{un_agenda}. Similarly, the Intergovernmental Panel on Climate Change (IPCC) Fifth Assessment Report Summary for Policymakers emphasizes that "indigenous, local and traditional knowledge systems and practices, including indigenous peoples' holistic view of community and environment, are a major resource for adapting to climate change, but these have not been used consistently in existing adaptation efforts" \cite{ipcc_ir5}. Traditional ecological knowledge (TEK) encapsulates indigenous cultural practices, world views, and ways of life which offer myriad epistemological and ontological approaches, including mythologies passed down as songs and stories, embedded in geographic representations, and more \cite{berkes2017sacred}. It is a field of study in anthropology defined as the cumulative body of knowledge, practices, and beliefs, passed down from one generation to the next. However, the romanticization of indigeneity poses a common trap in which the complexities of indigenous epistemologies are reduced and distorted. Historically, this has negatively impacted indigenous peoples and local communities \cite{povinelli2016geontologies}. 

A common perspective within the growing body of work on fairness, accountability, and transparency of AI is that various unwanted consequences of ML algorithms arise in some way from bias in the datasets used during the training and evaluation stages of model development, the limitations of the modeling techniques to capture real-world socio-technical contexts \cite{selbst2019fairness}, or as an artefact in the way people interact with a ML model after its deployment \cite{suresh2019framework}. For example, consider geographic diversity in image datasets. ImageNet is a widely-used image dataset consisting of 1.2 million labeled images, 45\% of which were taken in the United States, and the majority of the remaining images are from North America or Western Europe. Only 1\% and 2.1\% of the images come from China and India, respectively \cite{deng2009imagenet}. It has been shown that such representational biases lead to worse image classification performance for under-represented countries \cite{shankar2017no}. While ML methods typically rely on many iterations to reach a good solution, in many real-world applications there is often no possibility of having many feedback loops due to the high cost of failures. In the context of environmental sustainability unintended consequences of ML algorithms might have irreversible impacts. 

An interdisciplinary worldview can help practitioners recognize the need for multifaceted feedback loops in order to inform the discussion of how ML could meaningfully contribute to sustainability while adequately addressing issues of fairness, accountability, and transparency of how ML is used and what errors and edge-cases might occur. This could be facilitated by improving the capability of stakeholders to interface with and be accountable to one another. The concept of personhood for natural entities like rivers and forests, for example, could provide the ability for nature to interface with the other stakeholders within legal fictions. In 2017, the Whanganui River in New Zealand was legally recognized as a living being,  turning it into an agent whose voice is embodied by appointed guardians with the following duties: (1) to act and speak for and on behalf of the river; (2) to uphold the river’s recognition and values as an indivisible entity and as a legal person; (3) to promote and protect the environmental, social, cultural, and economic health and well-being of the river; (4) to take any other action reasonably necessary to achieve its purpose and perform its functions \cite{argyrou2019legal}.

The main contribution of our work is to demonstrate the need for data science and ML-based environmental projects to consider the diverse perspectives within place-based and traditional ecological knowledge systems. We provide an overview of how ML is currently used in the planning, execution, and monitoring stages of forest ecosystem restoration projects and aim to highlight how IPLCs could participate and immensely contribute to achieving positive environmental impacts. Ultimately, by building on existing climate governance models, we propose that new methodological and governance frameworks could reduce the negative impacts of the use of ML in restoration efforts.

\section{Literature Review}
There has been a growing interest in the intersection of data science, machine learning, and sustainability. Recent work by Rolnick et al. provides an overview of how ML is being applied to address climate change in the domains of: electricity systems, transportation, buildings and cities, industry, farms and forests, carbon sequestration, climate prediction, societal impacts, solar geoengineering, education, and finance \cite{rolnick2019tackling}. Related to ecosystem restoration projects, they highlight the use of computer vision techniques, transfer learning, reinforcement learning, and control theory \cite{rolnick2019tackling}. For example, computer vision on satellite images is used for modeling the amount of carbon stored in forests \cite{rodriguez2017quantifying}. The model is then used for predicting the carbon storage potential of deforested areas. ML is also employed for verification of conservation projects through satellite imagery \cite{santamaria2020truebranch}. More broadly, the field of Computational Sustainability has enabled computer scientists to apply their skills towards a broad range of sustainability challenges. Gomes et al. call for transformative synthesis by incorporating a combination of techniques from (1) data and ML, (2) optimization and dynamic models simulation, and (3) multi-agent crowdsourcing and citizen science \cite{gomes2019computational}. Levy et al. go on to investigate the links between specific areas of sustainability and broader social challenges by studying the role of global commodity supply chains in deforestation through ML modeling \cite{gardner2019transparency}. 

The question of malicious use and unintended consequences of ML in society has emerged from the early work of Norbert Wiener (1954) and others. More recently, such interest has led to the development of the areas of Social Informatics, Participatory Design, Computer-Supported Cooperative Work, Critical Algorithm Studies, and others. The main challenges they discuss include the negative impacts of biased and non-representative datasets \cite{mehrabi2019survey}, the lack of diversity and inclusion in the perspectives taken into account \cite{hoffmann2019fairness}, the need for algorithmic impact assessments that evaluate ML models based on the broader socio-technical context within which they are situated \cite{selbst2019fairness}. The decisions informing the design, development, and deployment of ML applications are often vulnerable to various forms of bias. The use of non-representative or culturally-skewed datasets, oversimplifications in optimization algorithms, model interpretation assumptions, and self-reinforcing feedback loops have lead to unintended and harmful consequences, for example reinforcement of gender, sex, and race inequities in facial recognition models \cite{mehrabi2019survey, angwin2019machine}. In the context of ecosystem restoration, biases may result in damages and harm to ecosystems and stakeholder communities. Principles for ML governance in regenerative ecosystem practices could make use of a framework proposed by Suresh and Guttag which identifies the following biases: (1) historical bias, (2) representation bias, (3) measurement bias, (4) aggregation bias, (5) evaluation bias, and (6) deployment bias \cite{suresh2019framework}. Margins of error can arise at each step of the ML process, highlighting the importance of a systemic approach for risk mitigation. Environmental and social problems are ultimately interconnected and therefore can only be overcome by changing fundamentally the way society is organized, and not simply through technical interventions \cite{rico1998gender}. Rico (1986) argues that "human and environmental dimensions of sustainability are inseparable, and that this link is a result both of the aggregate effect of social relationships and actions as they influence the natural ecology, and of the impact of environmental changes on society" \cite{rico1998gender}. Recent focus on investigating the downstream societal effects of the use of machine learning have lead to the development of multiple assessment and governance models \cite{assessmentList, latonero2018governing, gasser2017layered}. We build on these works and aim to highlight the need for such methodological frameworks in the way ML is used in restoration projects, also building on prior work in climate governance models.

The negotiations under the United Nations Framework Convention on Climate Change (UNFCCC) have instigated the 1997 Kyoto Protocol, the 2009 Copenhagen Accord, and the 2015 Paris Agreement. These can be considered as distinct climate governance models which have been studied by policymakers and other scholars. Held and Roger (2019) provide a comparative analysis of these models \cite{Held2018ThreeMO}, while other scholars bring political economy perspectives to the increasing but uneven uptake of transnational climate governance \cite{roger2017comparative}. Hale (2018) identify the unique characteristics of the Paris Agreement which allow for the re-framing of the international politics of climate change as a "catalytic" collective action problem \cite{hale2018catalytic}. In collective action problems, individual stakeholders could benefit from cooperation. Particularly, we hope our work could inspire new kinds of cooperation between applied machine learning researchers, policymakers, and IPLCs, leveraging traditional ecological knowledge systems. Uprety et al. provide a review of the use of TEK in ecological restoration and planning in 17 different kinds of restoration projects around the world \cite{uprety2012contribution}. However, scholars have also been concerned about the incorporation of traditional knowledge systems into "top-down" approaches to ecological restoration, pointing out ethical and social challenges \cite{chalmers2007expert}. We aim to bring these perspectives to data scientists and machine learning researchers working in sustainability, highlighting the need for participatory methodological frameworks that are centered on equity and collective action. 

\section{Leveraging Traditional Ecological Knowledge}

Ecologist Fikret Berkes (1999) brings together indigenous knowledge systems and Western scientific theories of conservation and biodiversity through his extensive fieldwork around the world. He explains TEK through four interrelated levels of ecosystem management, defined by a model known as the knowledge-practice-belief complex \cite{berkes2017sacred}. The levels in this model are: (1) the local knowledge of animals, plants, solids, and landscapes; (2) resource management, composed of local environmental knowledge, practices, and tools; (3) community and social organizing offering coordination, cooperation, and governance; and (4) worldviews involving general ethics and belief systems \cite{berkes2017sacred}. We propose that there's an opportunity to further bridge knowledge gaps between TEK and scientific knowledge through participatory methods building on decades of fieldwork. Utilizing methodologies such as participatory action research and appreciative inquiry in the planning stage of ecosystem restoration projects could improve the well-being of climate-vulnerable communities by building trust and facilitating collaborations among stakeholders. It could allow machine learning engineers, community members, and stakeholders to collectively make decisions about data collection, usage, storage, and removal protocols to ensure data sovereignty and build local capacity. We suggest that any ecosystem restoration effort needs to consider and plan for the economic and social-equity aspects of the proposed project. For example, ML projects might leverage IPLCs' deep knowledge and understanding of their land, while the IPLCs could co-benefit from the social and economic value created by the restored ecosystem services in the area. Luedeling et al. (2019) highlight the importance of drawing upon "expert knowledge to ensure that relevant constraints are considered" \cite{luedeling2019forest}. Working closely with experts and IPLCs will help guard against misguided strategy in the implementation stages by aligning incentives between stakeholders, including ecosystems, with well-being in mind.

There are numerous examples illustrative of TEK. The Maya Milpa is an agroecological system and multi-crop field historically employed in Latin America. These shifting cultivation systems, referred to as swiddens, comprise lands which are transformed for cultivation by means of skilled slashing and burning of vegetation. The Milpa woodland ecosystems are shaped without the use of fertilizers and pesticides, increasing soil fertility through forest succession \cite{nigh2013maya}. In the foothills Mount Kilimanjaro, the Chagga cultivate Kihamba forest gardens involving the management of multipurpose trees and shrubs in symbiosis with annual and perennial agricultural crops and livestock. The Kihamba are a densely vegetated tropical forest ecosystem resulting in high biodiversity with over 500 species in a single forest garden \cite{hemp2006banana}. In the Xingu River Basin of Amazonia, the Kayapó have been protecting one of the largest tropical forests on Earth. The Kayapó’s use of fire in agriculture successfully works as a soil fertilizer and stimulates forest regeneration. They construct forest islands, or apêtê, which maximize biodiversity utilizing a polycrop relay system. Resulting in 250 food plants and 650 medicinal plants within the forest gardens \cite{parker1992forest}. Tyson Yunkaporta emphasizes the importance of holistic reasoning to create dialogue between scientific and Indigenous knowledge systems. He elucidates five principles, each as symbols which are simultaneously further encoded into a singular symbolic image. This way of sensemaking through metaphors are comprised of story-mind, kinship-mind, dreaming-mind, ancestor-mind, and pattern-mind, which can be mapped onto the five fingers of a hand and made to interact. We suggest such processes could be incorporated into ML and data science approaches in the context of ecosystem regeneration \cite{yunkaporta2019sand}.

In what follows, we provide a brief analysis of how ML is used in the planning, executing, and monitoring stages of forest ecosystem restoration projects. We don't aim to provide a full overview of the ecological restoration practices and point interested readers to the work of Egan and Howell for a detailed review \cite{egan2005historical}.

\subsection{Planning of restoration}
ML on satellite and drone image data is being used in various ways throughout the planning stage of forest restoration projects. Through complex ML modeling, researchers aim to more accurately determine forest carbon sequestration potential worldwide \cite{carbonseq,rodriguez2017quantifying}. We propose that satellite and drone image data alone cannot be sufficient resources in the restoration planning stage because ML-based pattern recognition algorithms may fail to capture the significance of socio-ecological aspects of the land. In order to communicate the importance of these aspects, ethnographers have evolved the concept of cultural keystone place to denote "a given site or location with high cultural salience for one or more groups of people and which plays, or has played in the past, an exceptional role in a people's cultural identity, as reflected in their day to day living, food production and other resource-based activities, land and resource management, language, stories, history, and social and ceremonial practices" \cite{cuerrier2015cultural}. We suggest that ML practitioners will benefit from new organizing principles allowing for and incentivizing teaming up with people from interdisciplinary fields during all stages of an ecosystem restoration project. In order to partner with IPLCs, a technical ML team might employ qualitative research methods while collaborating with local municipalities, foresters, and other subject matter experts. The organizing principle needs to enable open and transparent participation of all involved stakeholders, including seeing the IPLCs as equally participating actors throughout the process.

Restoration efforts looking to enable higher levels of biodiversity need to assess the flow of ecosystem services in the target restoration areas \cite{cuerrier2015cultural}. Ecosystem services encompass supporting, provisioning, regulating, and cultural services \cite{cuerrier2015cultural}, where there exists a reciprocal relationship between people and ecosystems. Ethnographic surveys of communities impacted by ecological degradation could serve as helpful resources to ML practitioners. The consideration of cultural ecosystem services \cite{brancalion2014cultural} could further enable practitioners to plan how a proposed ML project could be integrated in a socio-economic and socio-ecological context. For example, the planning project stage could explore topics such as (1) impact on the well-being of local communities, (2) long-term environmental and biodiversity impact, (3) economic opportunities likely to be created through the implementation of the project, (4) provision of new educational and knowledge-generation value, and (5) resource stewardship.

Ecosystems maintain stability through interactions between various feedback loops \cite{neutel2017symmetry}. ML systems interfacing with ecosystems could create ecosystemic dependencies with ML feedback loops becoming an integral part of the ecosystem. We propose the assessment of necessary resources for ongoing ML efforts and the consideration of ML applications toward and beyond a threshold for ecosystemic self-sufficiency.

\subsection{Execution of restoration}
Environmental scholars have argued about the need for recognition of the roles of communities in actively cultivating, improving, and positively contributing to ecosystem services, challenging "the false concept of ecosystem services as a one-way flow of benefits from ecosystems to humans" \cite{comberti2015ecosystem}. Building on cultural ecosystem services as well as the concept of "services to ecosystems" \cite{comberti2015ecosystem}, we suggest ML practitioners who execute restoration projects need to see these projects as continuous reiterative efforts where involved stakeholders such as IPLCs might be the ones providing the needed services to ecosystems support in return for sharing the benefits from social, economic, and other values created by the restoration projects. We propose that in order to ensure equity in the IPLCs' involvement in restoration efforts, ML developers need to create rich interaction interfaces which lower the barriers to access, allow for people from varying backgrounds and levels of technical skills to participate, provide explanations for algorithmic outcomes \cite{doshi2017towards}, and allow for people to intervene at every step \cite{orseau2016safely}. 

\subsection{Monitoring of restoration}
Among the many applications of ML operating on satellite and drone image data in the monitoring stages of forest ecosystem restoration projects are detecting and mapping selective logging \cite{hethcoat2019machine}, high-resolution land cover mapping \cite{robinson2020human}, and predicting forest wildfire spread \cite{radke2019firecast}. Monitoring is also an important stage in the lifecycle of any ML model which is deployed in a real-world system. Continuous monitoring, integration, re-training and tuning of model parameters is part of a more general quality assurance process which ultimately aims to make sure that a ML model is able to perform efficiently as people are interacting with it. An increased number of feedback loops between the model-users and model-creators will positively contribute to models which are adaptive to the dynamic context within which they operate. As explored by Ortega et al., monitoring and control of system activity, framed as assurance, is an integral part of building safe artificial intelligence \cite{ortega2018building}. We suggest that community engagement could help large-scale continuous ML model assurance efforts, enabling the alignment of incentives between human and algorithmic actors. 

During the execution of the restoration stage, partners in the project collaboratively act and implement solutions to address an agreed-upon goal. Similarly, during the monitoring stage, the outcomes of their actions are collectively reviewed and evaluated. This could include engaging communities as a sensor network, sharing examples of the changes they see, or early indications that something new or different might happen. Should the efforts of the implementation be found unsuccessful or partially successful, the adjusted results are fed into the next iteration of the planning cycle. We suggest that throughout this adaptive process it is important to establish ongoing commitments for each project stakeholder. For the ML research team, this could include establishing practical ethical guidelines, adopting an impact assessment framework, involving impacted communities in active participation rather than passive acceptance to ensure cultural relevance to the community, as well as building community capacity and resiliency with new skills. This could also include an exit strategy of ML interventions once desired outcomes have been achieved.

\section{Enabling catalytic cooperation through a shift in ML governance models}
An interdisciplinary worldview helps us recognize the need for multifaceted feedback loops in order to inform the discussion of how ML could meaningfully contribute to Sustainability. As Gregory Bateson would say, some questions are not meant to be answered but they show us new perspectives about the relations between all involved actors \cite{bateson2000steps}. Similarly, the question of how to use ML to alleviate the fragility of our planet could bring awareness to these relations and inform new organizing principles. One such organizing principle is modeling the responsible design, development, and deployment of ML as a tragedy of the commons problem \cite{hardin2009tragedy,ostrom1990governing,reilly2001establishing,greco2004tragedy} as well as investigating how the problem of reducing the negative impacts of ML departs from the logic of the tragedy of the commons. Leveraging work on other tragedy of the commons systems such as climate action \cite{Held2018ThreeMO}, we could make more informed decisions about the organizing and governance principles that could enable positive results, reducing the negative impacts of ML-based ecosystem restoration efforts. 

The problem of reducing environmental degradation has the characteristics of joint products, preference heterogeneity, and increasing returns discussed by Hale \cite{hale2018catalytic}. Joint products means that actions could benefit the actors while also contributing public good to communities. Preference heterogeneity relates to the fact that there's no symmetry of preferences across actors and actions. In the tragedy of the commons model, the free-rider problem poses that an actor is generally disincentivized from action by the efforts of others. In reality, many actions reinforce themselves through a variety of feedback loops that generate increasing returns to action over time - action in the past can reduce the cost and increase the benefit of action in the future.

We propose that many ML applications fit the features of joint products, preference heterogeneity, and increasing returns, which creates the possibility for re-framing the methodological and governance frameworks towards a "catalytic cooperation" model \cite{hale2018catalytic}. An example of such a governance model is the 2015 Paris Agreement on climate change. By doing a comparative analysis of the problem structures of climate action and the use of ML, we find that they exhibit similarities in their distributional effects, the spread of individual vs. collective harms, and first-order vs. second order impacts. Hence, we propose that it is helpful to restructure and address ML governance questions through a catalytic cooperation model which recognizes that:

\begin{itemize}
  \item Good intentions are not good enough. Acknowledging the fallacy of technological solutionism, there's a need to stimulate incremental action within academia, the private sector, civil society, etc. Frameworks can facilitate measurability, which helps actionability and adds to the conceptual toolbelt for the assessment of complex problems arising from the interplay of many agents \cite{schiff2020ieee}.
  
  \item Scale matters. There's a need for new ways to participate. For every area of sustainability, it is possible to shift from a worldview where negative impacts of ML are diffused in society towards an increasing number of positive examples of ML contributing to socio-economic and socio-ecological well-being of people and the planet. Increasing the number of actors working on these issues lowers the costs and risks for more actors to become involved in this space until the kickstart of a "catalytic effect" resulting in a tipping point where the new behaviors and norms become self-reinforcing \cite{hale2018catalytic}.

    \item There's a need for intergenerational consent, consensus, commitments, and cooperation through transparent normative goal-setting and benchmarking. Metrics frameworks, standards, best-practices, and guidelines could contribute to an iterative evaluation process which enables collective action aligned with society's broader values and beliefs \cite{musikanski2018ieee}.

\end{itemize}

\section{Conclusion}
Ecological restoration and regeneration efforts sit at the heart of moving towards accelerating positive environmental impacts. However, there's a growing need for governance frameworks which could enable collective action through empowering community engagement, equity, and long-term impacts. Traditional ecological knowledge systems of indigenous peoples around the world have been globally recognized as a major asset in local restoration efforts. By bringing interdisciplinary perspectives to the work of data science and machine learning scholars, we aim to highlight the methodological opportunities and principles for integrating traditional ecological knowledge systems in the design, development, and deployment of ML-based forest restoration projects. We have provided an overview of how machine learning is used in the planning, execution, and monitoring stages of ecological restoration and hope to engage in applied work in our ongoing research. Future work needs to also consider what methodological frameworks could bridge the gaps between applied ML-based projects and environmental policy. We imagine that ensuring stakeholder equity could unleash conceptual tools, building on the principles we have hereby proposed. Furthermore, catalytic cooperation shows immense potential as a systemic approach towards such an integration between data scientists, indigenous and local communities, policymakers, and others, while optimizing for ecosystem regeneration, maximizing biodiversity, and community well-being. 

\bibliographystyle{ACM-Reference-Format}
\bibliography{main.bib}

\end{document}